\documentstyle[prd,aps,12pt]{revtex}

\newcommand{\be}{\begin{equation}}
\newcommand{\ee}{\end{equation}}
\newcommand{\bu}{{\bf u}}
\newcommand{\bv}{{\bf v}}
\newcommand{\bw}{{\bf w}}
\begin{document}
\draft
\preprint{Imperial/TP/94-95/46}
%@twocolumnfalse\endcsname

\title{Minimum Decision Cost for Quantum Ensembles}

\author{Dorje Brody
\footnote[1]{Electronic address: d.brody@tp.ph.ic.ac.uk}
and Bernhard Meister
\footnote[2]{Electronic address: b.mister@ic.ac.uk}
}

\address{* Blackett Laboratory, Imperial College,
South Kensington, London SW7 2BZ, U.K.}
\address{$\dagger$ Isac Newton Institute For Mathematical
Science, 20 Clarkson Road, Cambridge, CB3 0EH, U.K.}

\date{\today}

\maketitle

\begin{abstract}
For a given ensemble of $N$ independent and identically
prepared particles, we calculate the binary decision costs
of different strategies for measurement of polarised spin
1/2 particles. The result proves that, for any given values
of the prior probabilities and any number of constituent
particles, the cost for a combined measurement is always
less than or equal to that for any combination of separate
measurements upon sub-ensembles. The Bayes cost, which is
that associated with the optimal strategy (i.e., a
combined measurement) is obtained in a simple closed form.
\par
\end{abstract}


%\vskip2pc]

In a problem of experimental design, the task of the
experimentalist is to find an optimal observational strategy.
Ordinarily, one must choose among different strategies before
the data can be obtained, and hence one must perform a {\it
preposterior analysis}. When the experiment involves a
decision among different quantum mechanical states, such an
analysis is indeed important, since, unlike the classical case,
{\it virtual sampling}, i.e., repeated samplings of the same
system, are not generally permitted.  \par

There are a number of different approaches for finding an optimal
strategy. In the information-theoretic approach, one typically
determines the strategy that maximises the mutual information
(see, e.g., \cite{dv}), but this is generally difficult, owing to
the nonlinear nature of the Shannon information. In the minimax
approach \cite{ber}, one finds the strategy that minimises the
maximum cost (or loss) incurred by the decision among different
strategies. When certain {\it a priori} knowledge concerning the
nature of the state is available, then one may seek a strategy
that minimises the expected cost, using a Bayes procedure
\cite{ber}, \cite{deg}. \par

In the present Letter, we study the Bayesian approach to a
binary decision problem for an ensemble of polarised spin 1/2
particles. First, we briefly introduce the Bayesian approach to
quantum hypothesis testing. These notions, developed by
Helstrom and others (\cite{hel}, \cite{hol}, and \cite{ykl}),
are then applied to obtain the optimal strategy for a Bayes
decision between two quantum mechanical pure states, for an
ensemble of polarised spin 1/2 particles. In this example, we
first study the application of quantum Bayes sequential
analysis to the ensemble. The result is then compared with a
combined measurement of the entire ensemble, treated as a
single composite system. Other strategies consisting of
combined measurements of sub-ensembles are also considered.
The Bayes solution to the problem demonstrates that the Bayes
cost for separate sequential measurements of the individual
particles is the same as that of a combined measurement. This
result differs from that predicted by Peres and Wootters
\cite{pw}. Any other strategy turns out to entail a higher
expected cost. Nevertheless, we conclude, for the reasons given
below, that a combined measurement of the entire ensemble is,
in general, an optimal one.\par

First, consider a decision problem requiring a choice among
$M$ hypotheses $H_{1}, \cdots, H_{M}$ concerning a quantum
system. Hypothesis $H_{k}$ asserts that the density operator
of the system is $\hat{\rho}_{k}$, $(k = 1, \cdots, M)$, and
the prior probability of the $j$-th state is $\xi_{j}$, with
%%%%%%%%%%%%%%%%%%%%%%%%%%
\be
\sum_{k=1}^{M} \xi_{k}\ =\ 1\ .
\ee
%%%%%%%%%%%%%%%%%%%%%%%%%%
{}From past experience, one knows that the system is in the
$j$-th state with a relative frequency $\xi_{j}$. The
self-adjoint operators $\hat{\rho}_{k}$ act on the vectors of
a Hilbert space $\cal{H}$, are non-negative definite, and
have unit trace. \par
A {\it quantum decision strategy} is characterised by a
{\it probability operator measure} (pom) on ${\cal H}$, i.e.,
a set of $M$ non-negative definite self-adjoint operators
$\Pi_{j}$ satisfying
%%%%%%%%%%%%%%%%%%%%%%%%%%
\be
\sum_{j=1}^{M} \Pi_{j}\ =\ {\bf 1}\ .
\ee
%%%%%%%%%%%%%%%%%%%%%%%%%%
If this pom is applied to the system when hypothesis $H_{k}$
is true, then the conditional probability of choosing
hypothesis $H_{j}$ is given by
%%%%%%%%%%%%%%%%%%%%%%%%%%
\be
{\rm Pr}(X=j|W=k)\ =\  {\rm Tr}(\rho_{k}\Pi_{j})\ .
\ee
%%%%%%%%%%%%%%%%%%%%%%%%%%
Here, $X$ denotes the random variable that is to be observed,
and $W$, typically being the parameter, is the unknown state
of nature.\par
Now, let $C_{ij}$ be the cost of choosing hypothesis $H_{i}$
when $H_{j}$ is true. Then the expected cost of the
observational strategy specified by the pom $\{ \Pi_{j}\}$ is
\cite{hel}
%%%%%%%%%%%%%%%%%%%%%%%%%%
\be
{\bar C}\ =\ \sum_{i=1}^{M}\sum_{j=1}^{M} \xi_{j}C_{ij}
{\rm Tr}(\hat{\rho}_{j}\Pi_{i})\ \equiv\
{\rm Tr}\sum_{i}^{M} R_{i}\Pi_{i}\ ,
\ee
%%%%%%%%%%%%%%%%%%%%%%%%%%
where the Hermitian $risk$ $operators$ $R_{i}$ are defined by
%%%%%%%%%%%%%%%%%%%%%%%%%%
\begin{equation}
R_{i}\ =\ \sum_{j=1}^{M} \xi_{j}C_{ij}\hat{\rho}_{j}\ .
\end{equation}
%%%%%%%%%%%%%%%%%%%%%%%%%%
A set $\{ \Pi_{j}\}$ of pom that minimises the cost (4), under
the constraints (2), is defined as optimal and the cost is
Bayes, i.e., ${\bar C}={\bar C}^{*}$ (the suprescript * here
corresponds to the optimal strategy). Necessary and sufficient
conditions for the optimality of a pom are known to be
\cite{hol}, \cite{ykl} the self-adjointness of the operator
%%%%%%%%%%%%%%%%%%%%%%%%%%
\be
\Upsilon \ =\ \sum_{j=1}^{M}R_{j}\Pi_{j}\ =\
\sum_{j=1}^{M}\Pi_{j}R_{j}
\ee
%%%%%%%%%%%%%%%%%%%%%%%%%%
and the non-negative definiteness of the operator $R_{j} -
\Upsilon$ for all $j=1, \cdots, M$. The minimum expected Bayes
cost is thus
%%%%%%%%%%%%%%%%%%%%%%%%%%
\begin{equation}
{\bar C}^{*}(\xi,\{ \Pi_{j}^{*}\})\ =\ {\rm Tr}\ \Upsilon\ .
\end{equation}
%%%%%%%%%%%%%%%%%%%%%%%%%%
\par
In a simple case where $M=2$, i.e., for binary decisions, one
can easily verify \cite{hel} that the optimal pom is
projection valued, and the Bayes cost becomes
%%%%%%%%%%%%%%%%%%%%%%%%%%
\begin{eqnarray}
{\bar C}^{*}(\xi, \{ \Pi_{j}^{*}\})\ &=&\
\xi_{1}C_{11} + \xi_{2}C_{12} \nonumber \\
& &\ - \xi_{2}(C_{12}-C_{22})
\sum_{\eta_{i}>0} \eta_{i}\ ,
\end{eqnarray}
%%%%%%%%%%%%%%%%%%%%%%%%%%
where $\eta_{i}$ are the eigenvalues of the operator
$\hat{\rho}_{2} - \gamma \hat{\rho}_{1}$, with
%%%%%%%%%%%%%%%%%%%%%%%%%%
\be
\gamma\ =\ \frac{\xi_{1}(C_{21}-C_{11})}
{\xi_{2}(C_{12}-C_{22})}\ =\ \frac{\xi}{1-\xi}\ .
\ee
%%%%%%%%%%%%%%%%%%%%%%%%%%
Here and in the sequel, we choose a 0-1 cost structure; $C_{ij}
= 1-\delta_{ij}$, i.e., assign cost 1 to an incorrect decision
and 0 to a correct decision. Also, the prior probability for
state 1 is given by $\xi_{1}=\xi$, and hence $\xi_{2}=1-\xi$.
\par

Now, we consider an experiment where a physicist must
estimate (decide) the direction of polarisation of a given
ensemble of $N$ spin 1/2 particles, using a Stern-Gerlach
(s-g) device. The physicist knows that the particles have
been filtered through another s-g device with a magnetic
field in the $x-y$ plane at a constant angle $\theta_{1}$
or $\theta_{2}$ from the $x$-axis, and in either case the
spin up state has been selected. The physicist can select
the orientation angle $\phi$ of the detector relative to the
$x$-axis.  When the particle passes through the field of the
detector magnet, the physicist observes either the spin up
(head) or spin down (tail) state, whereupon he must decide
between the alternatives $\theta_{1}$ (i.e., the polarisation
direction $\theta = \theta_{1}$) and $\theta_{2}$. We do
not specify the values of the angles $\{ \theta_{k}\}$,
but the difference between the two angles is given by
$|\theta_{2}-\theta_{1}| = 2\delta$.\par

First, consider the case where the physicist performs
sequential observations of each individual spin 1/2 particle.
Suppose, for simplicity, that $N=1$. The physicist has to
decide, either before or after the observation, whether
the particle is polarised in the $\theta_{1}$ or
$\theta_{2}$ direction. If a decision were to be chosen
without any observation, then a Bayes decision against the prior
distribution $\xi(W)$ of $W$ (in this case, $W=1$ or $2$)
would be optimal. Suppose that $X$ (spin `up' or `down')
is observed before a decision is chosen. Then, the decision
process for the physicist follows the same procedure as the
previous case. However, the difference here is that the
distribution of $W$ has changed from the prior to the
posterior distribution. Hence, a Bayes decision against the
posterior distribution of $W$ is now optimal. \par

The conditional probability for observing the spin up
$(+1)$ state, when the state of the system is
$\hat{\rho}_{k}$, is given by
%%%%%%%%%%%%%%%%%%%%%%%%%%%%%%%%%%%%%%%%%%
\be
b_{k}(\phi)\ \equiv\ {\rm Pr}(X=+1|W=\theta_{k})\ =\
\cos^{2}(\frac{\theta_{k}-\phi}{2})\ .
\ee
%%%%%%%%%%%%%%%%%%%%%%%%%%%%%%%%%%%%%%%%%%
If one fixes the angle $\phi$, then the experiment is
entirely analogous to a classical coin tossing problem
\cite{bm}, with coins whose bias is given by the above $b_{k}$.
However, having the freedom to choose the angle $\phi$
for each value of the prior $\xi$, the physicist must
choose an optimal direction given by \cite{mh}
%%%%%%%%%%%%%%%%%%%%%%%%%%%%%%%%%%%%%%%%%%
\be
\phi_{opt}(\xi)\ =\ \tan^{-1}\left(
\frac{\xi\sin\theta_{1} - (1-\xi)\sin\theta_{2}}
{\xi\cos\theta_{1} - (1-\xi)\cos\theta_{2}}
\right) \ .
\ee
%%%%%%%%%%%%%%%%%%%%%%%%%%%%%%%%%%%%%%%%%%
Hence, we have a problem of tossing quantum coins whose
bias is a function of the prior probability $\xi$. \par

Having chosen the optimal angle $\phi_{opt}$, the
Bayes decision rule specifies that $\theta_{1}$
is to be chosen if the spin up state is observed, and
$\theta_{2}$ otherwise. The Bayes cost against the prior $\xi$,
when $N=1$, can easily be obtained by calculating the
eigenvalues of $\hat{\rho}_{2}-\gamma\hat{\rho}_{1}$,
with the result \cite{mh}
%%%%%%%%%%%%%%%%%%%%%%%%%%%%%%%%%%%%%%%%%%
\be
{\bar C}^{*}(\xi,1)\ =\ \frac{1}{2}
\left( 1 - \sqrt{2\xi^{2} - (2 + \cos2\delta)\xi
+ 1}\ \right)\ .
\ee
%%%%%%%%%%%%%%%%%%%%%%%%%%%%%%%%%%%%%%%%%%
\par
Now, suppose that $N=2$, and the result of measurement of
the first particle has been obtained. As mentioned above,
the physicist must follow the same procedures as in the case
$N=1$, with the posterior distribution $\xi(\pm)$ instead
of the prior $\xi$. From Bayes' theorem, the posterior
probability that $\theta=\theta_{1}$ is given by
%%%%%%%%%%%%%%%%%%%%%%%%%%%%%%%%%%%%%%%%%%
\be
\xi(+)\ =\ \frac{b_{1}(\phi)\cdot\xi}
{b_{1}(\phi)\cdot\xi + b_{2}(\phi)\cdot(1-\xi)}\
\ee
%%%%%%%%%%%%%%%%%%%%%%%%%%%%%%%%%%%%%%%%%%
or
%%%%%%%%%%%%%%%%%%%%%%%%%%%%%%%%%%%%%%%%%%
\be
\xi(-)\ =\ \frac{(1-b_{1}(\phi))\cdot\xi}
{(1-b_{1}(\phi))\cdot\xi + (1-b_{2}(\phi))\cdot(1-\xi)}\ ,
\ee
%%%%%%%%%%%%%%%%%%%%%%%%%%%%%%%%%%%%%%%%%%%
according to the outcome ($+$ or $-$) of the first
measurement. The optimal orientation angle, before performing
the second measurement, is now given by
$\phi_{opt}(\xi(+))$ or $\phi_{opt}(\xi(-))$, accordingly.
The Bayes cost for this case ($N=2$) is given by the
weighted average, i.e.,
%%%%%%%%%%%%%%%%%%%%%%%%%%
\[
{\bar C}^{*}\ =\  b_{1} \xi {\bar C}^{*}(\xi(+),1) +
b_{2} (1-\xi) {\bar C}^{*}(\xi(-),1)\ .
\]
%%%%%%%%%%%%%%%%%%%%%%%%%%
\par
Next, we consider an arbitrary number $N$ of particles.
Again, the procedures are the same as above,
except that the prior is now replaced by one of the
$2^{N-1}$ posteriors [$\xi(++\cdots +)$, $\cdots$], after
observations of $N-1$ particles. In a classical
Bayes decision procedure \cite{ber}, \cite{deg}, it is
difficult (or impossible) to obtain the Bayes cost as a
closed function of $N$. The reason is that, first, one must
study the {\it tree} \cite{lin} of the posterior
distributions, with branches proliferating as $\sim 2^{N}$. To
each branch (i.e., posterior) of the tree, one associates the
cost ${\bar C}^{*}(\cdot, 1)$, and then calculates the weight
(probability) for the sequence of outcomes associated with that
branch. After these considerations, one can, in principle,
obtain the weighted average of the cost, which involves
$2^{N-1}$ terms. (Note that, for classical coins, the branches
of the posterior tree do recombine and hence proliferate as
$\sim N$. However, the weights associated with the branches do
not recombine, and therefore one cannot avoid the consideration
of $2^{N-1}$ terms.)\par

In the case of our ``quantum coins'', the situation appears
even worse, since, after each observation, the physicist
must turn the device in accordance with formula (11). This
results in changing the bias $b_{k}(\phi)$ of the ``coins''
at each stage, and hence one must also incorporate the bias tree
(which proliferates $\sim 2^{N}$). However, it turns out
that this optimal orientation forces the posterior tree to
recombine into two branches, i.e.,
%%%%%%%%%%%%%%%%%%%%%%%%%
\be
\xi(n,\pm)\ =\ \frac{1}{2} \left( 1 \pm
\sqrt{ 1 - 4\xi(1-\xi)\cos^{2(n-1)}\delta}\ \right)\ ,
\ee
%%%%%%%%%%%%%%%%%%%%%%%%%
where $\pm$ corresponds to the outcome of the last
($n-1$-th) trial being spin up $(+)$ or down $(-)$. This
result can be proven by induction as follows. First, for $n=1$,
it is easily verified that $\xi(1,\pm)=\xi(\pm)$ as given in (13)
and (14). Next, assume that the last ($n-1$-th) outcome of the
trial is $(-)$, and that the posterior is given by the above
$\xi(n,-)$. Then, if the next trial outcome is $(+)$, follows
from Bayes' theorem, that the posterior distribution, after
$n+1$ observations, is given by
%%%%%%%%%%%%%%%%%%%%%%
\[
\xi(\cdots -+)\ =\ \frac{b_{1}(\phi)\cdot\xi(n,-)}
{b_{1}(\phi)\cdot\xi(n,-) + b_{2}(\phi)
\cdot(1-\xi(n,-))}\ ,
\]
with $\phi = \phi_{opt}(\xi(n,-))$. After some algebra, one
can show that the above $\xi(\cdots -+) = \xi(n+1,+)$. The other
three cases [$\xi(\cdots --)$, etc.] can also be treated in
the same manner. \par

Although the weights for different branches neither recombine
in the quantum case, since ${\bar C}^{*}(\xi(n,+),1) =
{\bar C}^{*}(\xi(n,-),1)$, the final average cost is just
${\bar C}^{*}(\xi(n,\pm),1)$ times the sum of all the
different weights (which is just 1), and hence we finally
deduce that the Bayes cost for sequential observations is
%%%%%%%%%%%%%%%%%%%%%%%%%%%%%%%%%%%%%%%%%%%
\begin{eqnarray}
{\bar C}^{*}(\xi,N)\ &=&\ {\bar C}^{*}(\xi(N-1,\pm), 1)
\nonumber \\
&=&\ \frac{1}{2}\left( 1 - \sqrt{ 1 - 4\xi(1-\xi)
\cos^{2N}\delta } \ \right)\ ,
\end{eqnarray}
%%%%%%%%%%%%%%%%%%%%%%%%%%%%%%%%%%%%%%%%%%%
for either value of the $N-1$-th outcome ($+$ or $-$).\par

Next, consider the case where the physicist treats the
entire ensemble as a single composite system. The total
spin of a system with $N$ particles is just $N/2$, and
the density operator for a spin $N/2$ particle polarised
in the direction ${\bf n} = (\cos\theta, \sin\theta, 0)$
is given by
%%%%%%%%%%%%%%%%%%%%%%%%%%%%%%%%%%%%%%%%%%
\be
\left(\hat{\rho}(\theta)\right)_{mn}\ =\ 2^{-N}
\sqrt{ _{N}C_{m} \ _{N}C_{n}}
e^{-i(m-n)\theta} \ ,
\ee
%%%%%%%%%%%%%%%%%%%%%%%%%%%%%%%%%%%%%%%%%%
where $(n, m) = 0, \cdots, N$. According to the result in (8),
one must find the
eigenvalues of the matrix $\hat{\rho}_{2}-\gamma
\hat{\rho}_{1}$ in order to obtain the Bayes cost.
We first show that the matrix $\hat{\rho}_{2}-\gamma
\hat{\rho}_{1}$ is of rank two, and thus has only two non-zero
eigenvalues. Define two vectors ${\bf u} = \{ u_{n}\}$ and
${\bf v} = \{ v_{n}\}$ by
%%%%%%%%%%%%%%%%%%%%%%%%%%%%%%%%%%%%%%%%%%
\be
u_{n}\ \equiv\ 2^{-N/2} \sqrt{ _{N}C_{n}}
e^{in\theta_{1}}\ ,
\ee
and
\be
v_{n}\ \equiv\ 2^{-N/2} \sqrt{ _{N}C_{n}}
e^{in\theta_{2}}\ .
\ee
%%%%%%%%%%%%%%%%%%%%%%%%%%%%%%%%%%%%%%%%%%
Then, $(\hat{\rho}_{1})_{mn} = u^{*}_{m}u_{n}$ and
$(\hat{\rho}_{2})_{mn} = v^{*}_{m}v_{n}$. Since the
inner product $\bu \cdot \bu^{*} =
\bv \cdot \bv^{*} = 1$, one obtains
%%%%%%%%%%%%%%%%%%%%
\[
{\hat \rho}_{1}\bu^{*} \ =\
\sum_{n}(\hat{\rho}_{1})_{mn}u^{*}_{n} \ =\ \bu^{*}
\]
and similarly, $\hat{\rho}_{2}\bv^{*} = \bv^{*}$. Now, let
$\bw$ and $\lambda$ be an eigenvector and the corresponding
eigenvalue of the matrix
$\hat{\rho}_{2}-\gamma\hat{\rho}_{1}$, i.e.,
%%%%%%%%%%%%%%%%%%%%%%%%%%%%%%%%%%%%%%%%%%%
\be
(\hat{\rho}_{1}-\gamma\hat{\rho}_{2}) \bw \ =\
\lambda \bw\ .
\ee
%%%%%%%%%%%%%%%%%%%%%%%%%%%%%%%%%%%%%%%%%%%
We may expand the eigenvector $\bw$ in terms of a basis
that contains either $\bu^{*}$ or $\bv^{*}$, i.e.,
$\bw = c_{1}\bu^{*} +  \bu^{*}_{\perp}$ or
$\bw = c_{2}\bv^{*} +  \bv^{*}_{\perp}$. Here, $\bu^{*}_{\perp}$
denotes some vector orthogonal to $\bu^{*}$, and similarly for
$\bv^{*}_{\perp}$. However, since $\hat{\rho}_{1}
\bu^{*}_{\perp} = \hat{\rho}_{2}\bv^{*}_{\perp} = 0$, we have
%%%%%%%%%%%%%%%%%%%%%%%%%%%%%%%%%%%%%%%%%%%
\be
\lambda \bw\ =\ c_{1}\bu^{*} - \gamma c_{2}\bv^{*} \ .
\ee
%%%%%%%%%%%%%%%%%%%%%%%%%%%%%%%%%%%%%%%%%%%
Therefore, the matrix $\hat{\rho}_{2}-\gamma\hat{\rho}_{1}$
is of rank two, as claimed. On the other hand, if we form the
inner product of the two vectors $\bw = c_{1}\bu^{*} +
\bu^{*}_{\perp}$ and $\bu$, we obtain
%%%%%%%%%%%%%%%%%%%%%%%%%%%%%%%%%%%%%%%%%%%
\be
\bw\cdot \bu\ =\ c_{1}\ =\ \frac{c_{1}}{\lambda}
- \frac{\gamma}{\lambda}c_{2}(\bv^{*}\cdot \bu)\ ,
\ee
and similarly,
\be
\bw\cdot \bv\ =\ c_{2}\ =\ \frac{c_{1}}{\lambda}
(\bu^{*}\cdot \bv) - \frac{\gamma}{\lambda}c_{2}\ .
\ee
%%%%%%%%%%%%%%%%%%%%%%%%%%%%%%%%%%%%%%%%%%%
Without any loss of generality, we may now set
$c_{1}=1$, and then by eliminating $c_{2}$ from the
above equations, we obtain the eigenvalues of the
matrix $\hat{\rho}_{2}-\gamma\hat{\rho}_{1}$, i. e.,
%%%%%%%%%%%%%%%%%%%%%%%%%%%%%%%%%%%%%%%%%%%
\be
\lambda_{\pm}\ =\ \frac{1}{2}
\left\{ (1 - \gamma) \pm \sqrt{
(1-\gamma)^{2} - 4\gamma(\Delta^{2}-1)} \right\}\ ,
\ee
%%%%%%%%%%%%%%%%%%%%%%%%%%%%%%%%%%%%%%%%%%%
where
%%%%%%%%%%%%%%%%%%%%%%%%%%%%%%%%%%%%%%%%%%%
\begin{eqnarray}
\Delta^{2}\ &=&\ (\bv^{*}\cdot \bu)
(\bu^{*}\cdot \bv) \nonumber \\
&=&\ \left| 2^{-N}
\sum_{m=0}^{N}\ _{N}C_{m}
e^{2im\delta} \right| ^{2} \ =\
\cos^{2N}(\delta)\ .
\end{eqnarray}
%%%%%%%%%%%%%%%%%%%%%%%%%%%%%%%%%%%%%%%%%%%
Therefore, the binary Bayes decision cost for a spin $N/2$
particle is
%%%%%%%%%%%%%%%%%%%%%%%%%%%%%%%%%%%%%%%%%%%
\be
{\bar C}^{*}(\xi,N)\ =\ \frac{1}{2}\left(
1 - \sqrt{ 1 - 4 \xi (1-\xi) \cos^{2N}\delta
}\ \right) \ .
\ee
%%%%%%%%%%%%%%%%%%%%%%%%%%%%%%%%%%%%%%%%%%%
One immediately observes that the above cost (26) is the
same as that obtained from sequential analysis, given by (16).
Hence, the Bayes solution to our optimisation problem states
that a combined measurement is as advantageous as sequential
measurements. These two strategies, however, are not the only
ones, and many other partially combined measurement procedures
are possible. However, in the present formalism of sequential
analysis, the only effect of any intermediate measurements,
either partially combined or not, consists in updating the
posterior distributions. Since the Bayes cost is a monotonically
decreasing function of the number of updating steps, this implies
that any partially combined measurements will increase the cost.
Therefore, we may now conclude that the optimal measurement
strategy consists in either performing a combined measurement of
the entire ensemble or performing sequential measurements of the
individual particles. Any other strategies will result in higher
costs. \par

This result is quite different from that expected
by Peres and Wootters, who conjectured that sequential
measurements can never be as efficient as a combined
measurement \cite{pw}. However, it is important to note
that their conjecture is based upon an information-theoretic
approach, and the solution of an optimisation problem using
a Bayesian approach can yield a different result. Massar and
Popescu \cite{mp}, on the other hand, have proved the above
mentioned conjecture explicitly for the case $N=2$. The
method used therein is effectively similar to a Bayesian
approach, without the use of the prior distributions. However,
when a prior distribution is available, the Bayes solution is
known to be optimal in general \cite{ber}. If prior knowledge
is not available, one can still apply the Bayesian approach,
using a non-informative prior. The analysis of such cases is,
however, beyond the scope of the present Letter. \par

Throughout the present Letter, we have only considered the cost
associated with making decisions. In any practical situation,
on the other hand, one must take into consideration other
costs (e.g., the observational cost, the cost of analysing
the results, etc.). In our example of sequential analysis, for
example, at each stage before performing an observation, the
physicist must analyse the previous results in order to
determine the optimal turning angle. One might argue that
\cite{ber} the analysing cost can be ignored, since, after all,
scientists are so underpaid that the cost of their labors is
usually negligible! Nonetheless, the observational costs cannot
be ignored in general. Assuming the linearity of the utility
function (e.g., that the total cost is just the sum of the
decision cost and the observational costs), it is clear that any
separate measurements will result in a higher total cost, since
the decision cost for optimal sequential measurements (i.e.,
sequential measurements with optimal angular orientations) can
never be lower than that for a combined measurement. Therefore,
we conclude, after these considerations, that a combined
measurement is optimal in general.\par

In connection with the decision problem for classical coins
which was briefly mentioned above, it is interesting to note that
all the quantum results obtained by calculating the eigenvalues
of the density operators can, in principle, be recovered from
purely classical calculations, even for sequential measurements,
if and only if the spins of the particles concerned are 1/2. That
is, provided one does not perform any combined measurements, the
results can be obtained from classical calculations. More details
of this, as well as a treatment including the observational costs,
may be found in \cite{bm}. (See, also \cite{mh} for a comparison
between classical and quantum coin tossings.)\par

The authors acknowledge their gratitude to J. T. Key,
and J. D. Malley for useful discussions of the foregoing
topics. \par

\end{document}